# Magnetic critical behavior of the van der Waals $Fe_5GeTe_2$ crystal with near room temperature ferromagnetism

Zhengxian Li[1,2,3], Wei Xia[1,2,3], Hao Su[1,2,3], Zhenhai Yu[1], Yunpeng Fu[1], Leiming Chen[4], Xia Wang[1,5], Na Yu[1,5], Zhiqiang Zou[1,5], Yanfeng Guo[1*]

[1]School of Physical Science and Technology, ShanghaiTech University, Shanghai 201210, China

[2]Shanghai Institute of Optics and Fine Mechanics, Chinese Academy of Sciences, Shanghai 201800, China

[3]University of Chinese Academy of Sciences, Beijing 100049, China

[4]School of materials science and engineering, Henan key laboratory of aeronautic materials and application technology, Zhengzhou University of Aeronautics, Zhengzhou, Henan, 450046

[5]Analytical Instrumentation Center, School of Physical Science and Technology, ShanghaiTech University, Shanghai 201210, China

[*]Corresponding author:

guoyf@shanghaitech.edu.cn.

The van der Waals ferromagnet $Fe_5GeTe_2$ has a Curie temperature $T_C$ of about 270 K, which can be raised above room temperature by tuning the Fe deficiency content. To achieve insights into its ferromagnetic exchange, we have studied the critical behavior by measuring the magnetization in bulk $Fe_5GeTe_2$ crystal around the ferromagnetic to paramagnetic phase transition. The analysis of the magnetization by employing various techniques including the modified Arrott plot, Kouvel-Fisher plot and critical isotherm analysis achieved a set of reliable critical exponents with $T_C$ = 273.7 K, $\beta$ = 0.3457 ± 0.001, $\gamma$ = 1.40617 ± 0.003, and $\delta$ = 5.021 ± 0.001, suggesting a three-dimensional magnetic exchange with the distance decaying as $J(r) \approx r^{-4.916}$, which is close to that of a three-dimensional Heisenberg model with long-range magnetic coupling.

# I. INTRODUCTION

A prominent virtue of the quasi-two-dimensional (2D) van der Waals (vdW) bonded materials is that they could be exfoliated into multi- or single layer, thus making them useful in various novel heterostructures and devices. Moreover, the vdW materials in the 2D limit exhibit extraordinary physical properties, such as those observed in the intensively studied graphene and transition metal dichalcogenides [1-6], etc. Known as the Merin-Wagner theorem [7], intrinsic long-range magnetic order can not appear in the isotropic magnetic 2D limit because the strong thermal fluctuations in such case prohibit the spontaneous symmetry breaking and hence the long-range magnetic ordering. Nevertheless, a small anisotropy is sufficient to open up a sizable gap in the magnon spectra and consequently stabilizes the magnetic order against finite temperature. This picture has been realized by the observation of long-range ferromagnetic (FM) order in mono- or few-layer $CrI_3$ [8], $Cr_2Ge_2Te_6$ [9], $Cr_2Si_2Te_6$ [10], $VSe_2$ [11], and $MnSe_2$ [12], etc. The vdW magnets in the 2D limit host rich magneto-electrical, magneto-optical, or spin-lattice coupling effects that are capable of exhibiting intriguing properties which are scarcely observed in bulk. Very recently, current-induced magnetic switch was observed in the few-layer $Fe_3GeTe_2$ [13], demonstrating the vdW magnets a versatile platform for nanoelctronics. Moreover, heterostructures constructed by using vdW magnets have profound valleytronics and spintronics device applications [14, 15]. For example, the tunneling magnetoresistance (MR) in spin-filter magnetic vdW $CrI_3$ heterostructures even approaches $1.9 \times 10^4$%, remarkably superior to that constructed by using conventional magnetic thin films [16]. The easy exfoliation, weak interlayer coupling, and tunability of magnetic properties make the vdW magnets a model family of materials for exploring exotic phenomena and finding novel applications.

In the handful FM vdW magnets, the physical properties in the 2D limit differ from each other due to rather complex magnetic interactions. The semiconducting monolayer $CrI_3$ is an Ising ferromagnet with very low Curie temperature ($T_C$) of about 45 K due to the weak superexchange interaction along the Cr-I-Cr pathway [8, 17].

The similar weak FM superexchange in the Heisenberg magnet bilayer $Cr_2Ge_2Te_3$ also results in a low $T_C$ of ~ 30 K, and FM order is even not present in the monolayer [9]. As a contrast, the FM exchange with an itinerant character mediated by carriers in metallic $Fe_3GeTe_2$ monolayer is much stronger than the superexchange in $CrI_3$ and $Cr_2Ge_2Te_6$, thus yielding a remarkably higher $T_C$ of about ~ 130 K, which can be raised even above room temperature by using the ionic gating technique [18, 19].

The tremendous efforts in perusing high $T_C$ more recently led to the discovery of a $T_C$ of ~ 260 -310 K in the bulk quasi-2D vdW $Fe_5GeTe_2$ and ~ 280 - 350 K in exfoliated thin flakes [20, 21]. Interestingly, similar as $Fe_{3-x}GeTe_2$, bulk $Fe_{5-x}GeTe_2$ shows a tunable $T_C$ ranging from ~ 130 K to ~ 230 K by controlling the Fe deficiency content $x$, suggesting the detrimental role of Fe in the magnetic exchange. A reversible magnetoelastic coupled first-order transition near 100 K was detected by neutron powder diffraction [20]. Considering the exotic physical properties in exfoliated $Fe_3GeTe_2$ nanoflakes and its heterostructures, such as the extremely large anomalous Hall effect [22], planar topological Hall effect [23], Kondo lattice physics [24], anisotropic magnetostriction effect [25], spin filtered tunneling effect [16], magnetic skyrmions [26], etc., $Fe_5GeTe_2$ would also be expected to provide extraordinary opportunities to explore intriguing physical properties. To well understand the physical properties of $Fe_5GeTe_2$, the magnetic exchange model should be established first. However, the direct measurements on the magnetic structure are absent yet. Alternatively, study on the magnetic critical behavior and analysis of the critical exponents in vicinity of the paramagnetic (PM) to FM transition region could yield valuable insights into the magnetic exchange and properties. For example, the method has established the magnetic exchange models for $CrI_3$ [27], $VI_3$ [28], $Fe_3GeTe_2$ [29, 30], $Co_2TiSe$ [31], and $Fe_{0.26}TaS_2$ [32], etc. In this work, we have reported the investigation on the critical behavior of $Fe_5GeTe_2$, which finds that the obtained set of critical exponents are close to those calculated from the renormalization group approach for a long-range 3D Heisenberg model with the magnetic exchange distance decaying as $J(r) \approx r^{-4.916}$.

## II. EXPERIMENTAL

Single crystals were grown from chemical vapor transport (CVT) technique by using iodine as the transport agent, similar as the method described in [20, 21]. The crystallographic phase and crystal quality were examined on a Bruker D8 single crystal X-ray diffractometer (SXRD) with Mo $K_\alpha$ ($\lambda$ = 0.71073 Å) at 300 K. The chemical compositions and uniformity of stoichiometry were checked by the energy dispersive spectroscopy (EDS) at several spots on the crystals. The direct current (dc) magnetization was measured on the Quantum Design magnetic properties measurement system (MPMS-3) with the magnetic field applied parallel to *c*-axis of the crystal. Isothermal magnetizations were collected at a temperature interval of 1 K in the temperature range of 261- 285 K, which is just around $T_C$ (~ 270 K). It should be noted that each curve was initially magnetized. The applied magnetic field was corrected by considering the demagnetization factor, which was used for the analysis of critical behavior.

## III. RESULTS AND DISCUSSION

The SXRD measurement gives the space group $R\bar{3}m$ (No.166) of the crystal structure with lattice parameters *a* = 4.05(1) Å, *c* = 29.23(2) Å, and *α* = *β* = 90°, *γ* = 120°, consistent with previously reported values in [20] but slightly larger than those in [21], presumably due to the smaller Fe deficiency content in our crystals, which is determined as *x* ~ 0.09 by the EDS chracterizations. We thereafter still use $Fe_5GeTe_2$ to represent the used crystals in this experiment. The crystal structure, schematically shown in Figs. 1(a)-(b) seen from the *a*- and *c*-axis respectively, is drawn based on analyzing the SXRD data. The structure of $Fe_5GeTe_2$, analogue to $Fe_3GeTe_2$, is basically built up by 2D slabs of Fe and Ge between the van der Waals gapped Te layers [20, 33]. The perfect reciprocal space lattice of SXRD without any other miscellaneous points, seen in Figs. 1(d) - (f), indicates pure phase and high quality of the crystal. Previous studies unveiled that $Fe_5GeTe_2$ has two similar crystal structures when the synthesis methods are different [21]. One structure has a higher symmetry in a space group $R\bar{3}m$ and the other one has the space group of *R3m*. These structures both are with rhombohedral lattice centering three $Fe_5GeTe_2$ layers in each unit cell. The structure with a higher symmetry contains three Fe sites in each unit cell, where the Fe(1) site, marked in Fig. 1(a), is treated as a split site occupying either above or below the neighboring Ge site. As a contrast, in the lower symmetry model the equivalent Fe(1) site is never treated as a split site and is always "up" in a layer. Analogue to $Fe_{3-x}GeTe_2$, the equivalent Fe(1) site could be vacant, thus allowing the Fe deficiency in both compounds to tune the $T_C$.

Fig. 2(a) depicts the temperature dependence of magnetization *M*(*T*) for $Fe_5GeTe_2$ measured with zero-field-cooling (ZFC) and field-cooling (FC) mode under the applied magnetic field *H* = 1 kOe along the *ab*-plane of the crystal. The magnetization displays an abrupt PM to FM transition at ~ 270 K and no clear separation between ZFC and FC curves. The inset of Fig. 2(a) is the inverse temperature dependent magnetic susceptibility $\chi^{-1}(T)$ with the dotted straight line representing the Curie-Weiss law fitting. It can be seen that $\chi^{-1}(T)$ deviates from the straight line near 295 K which is much higher than $T_C$. The obtained Weiss temperature is 283 K which is also higher than $T_C$, indicating a strong FM interaction. The Curie-Weiss fitting also gives the effective moment as $\mu_{eff}$ = 6.659 $\mu_B$/Fe. Considering the varied effective magnetic moment of $Fe^{2+}$ with the values raging from 4.90 to 6.70 $\mu_B$ in various materials including sphalerite and monoclinic pyroxenes obtained from magnetic susceptibility analysis [34] and the Fe deficiency in our crystals, the value we obtained from the Curie-Weiss law fitting is reasonable. The FM ground state can also be demonstrated by the isothermal magnetization *M*(*H*) shown in Fig. 2(b) measured at 2 K. The clear low coercive field indicates a soft ferromagnetism in $Fe_5GeTe_2$, which is similar as that of $Fe_3GeTe_2$ [29, 30]. The observed saturation magnetization $M_S$ along the *c*-axis was 111 emu/g with a moment of 2.4 $\mu_B$/Fe. The initial isothermal magnetizations in the temperature range of 261 − 285 K measured with *H*//*c*-axis were shown in Fig. 2(c) and the Arrott plot [35], that is, $M^2$ vs *H*//*M*, is shown in Fig. 2(d). The positive slope of all $M^2$ vs *H*//*M* curves, according to the Banerjee's criterion [36], indicates that the PM to FM transition has a second-order in nature. By using the Arrott plot, the mean Landau mean-field theory with the critical exponents $\beta = 0.5$ and $\gamma = 1.0$ is involved, thus the $M^2$ vs $H/\!/M$ curves should be straight and parallel to each other in the high magnetic field region. Additionally, the isothermal magnetization at $T_C$ should pass through the origin. However, seen in Fig. 2(d), $M^2$ vs $H/\!/M$ curves are nonlinear with exhibiting a downward curvature, suggesting that the Landau mean-field theory is not applicable to $Fe_5GeTe_2$. The failure of the Arrott plot within the framework of Landau mean-field theory lies in that the itinerant ferromagnetism in $Fe_5GeTe_2$ should have significant electronic correlations and spin fluctuations, which however are neglected in the Landau mean-field theory.

The second-order PM to FM phase transition in $Fe_5GeTe_2$ can be described by the magnetic equation of state and is characterized by critical exponents $\beta$, $\gamma$ and $\delta$ that are mutually related. According to the scaling hypothesis, for a second-order phase transition, the spontaneous magnetization $M_S(T)$ below $T_C$, the inverse initial susceptibility $\chi_0^{-1}(T)$ above $T_C$ and the magnetization $M$ at $T_C$ can be used to obtain $\beta$, $\gamma$ and $\delta$ from the equations [37]:

$$M_S(T) = M_0(-\varepsilon)^\beta, \varepsilon < 0, T < T_C, \qquad (1)$$

$$\chi_0^{-1}(T) = (h_0/m_0)\varepsilon^\gamma, \varepsilon > 0, T > T_C, \qquad (2)$$

$$\text{and } M = DH^{1/\delta}, \varepsilon = 0, T = T_C, \qquad (3)$$

where $\varepsilon = (T - T_C)$ is the reduced temperature, and $M_0$, $h_0/m_0$, and $D$ are the critical amplitudes. Though the Landau mean-field theory can not be used, the critical isothermal magnetizations, alternatively, can be analyzed with the Arrott-Noakes equation of state [38]:

$$(H/M)^{1/\gamma} = a\varepsilon + bM^{1/\beta}, \qquad (4)$$

where $a$ and $b$ are fitting constants. Five different models including the 2D Ising model ($\beta = 0.125$, $\gamma = 1.75$) [39], the 3D Heisenberg model ($\beta = 0.365$, $\gamma = 1.386$) [39], the 3D Ising model ($\beta$ = 0.325, $\gamma$ = 1.24) [39], the 3D-XY model ($\beta$ = 0.345, $\gamma$ = 1.316) [40] and the tricritical mean-field model ($\beta$ = 0.25, $\gamma$ = 1.0) [41] were used for the modified Arrott plots, which are shown in Figs. 3(a)-(e). One can see that the lines in Figs. 3(d) and (e) are not parallel to each other, suggesting that the tricritical mean-field model and 2D Ising model are not appropriate to $Fe_5GeTe_2$. In Figs. 3(a)-(c), all lines in each figure are almost parallel to each other in the high magnetic field region, thus making the choice of an appropriate model for $Fe_5GeTe_2$ impossible in this step. As we mentioned above, the modified Arrott plot should be a set of parallel lines in the high magnetic field region with the same slope of $S(T) = dM^{1/\beta}/d(H/M)^{1/\gamma}$. The normalized slope $NS$ is defined by $NS = S(T)/S(T_C)$, which enables us an easy comparison of the $NS$ of different models and to select out the most appropriate one with the ideal value of unity. The $NS$ values versus the temperature for different models are plotted in Fig. 3(f), which clearly show that the $NS$ of the 2D Ising model has the largest deviation from unity. One can see that when $T > T_C$, $NS$ of the 3D Ising model is close to unity, while when $T < T_C$ the 3D XY model seems as the best. This indicates that the critical behavior of $Fe_5GeTe_2$ may not belong to a single universality class. The fact also indicates that the magnetic character of $Fe_5GeTe_2$ is nearly isotropic above $T_C$ and the enhancement of the anisotropic exchange below $T_C$.

In order to achieve in-depth insights into the nature of the PM to FM transition in $Fe_5GeTe_2$, the precise critical exponents and critical temperature should be obtained. In the modified Arrott plot, the linear extrapolation of the nearly straight curves from the high magnetic field region intercepting the $M^{1/\beta}$ and $(H/M)^{1/\gamma}$ axes yields reliable values of $M_S(T)$ and $\chi_0^{-1}(T)$, respectively. The extracted $M_S(T)$ and $\chi_0^{-1}(T)$ can be used to fit the $\beta$ and $\gamma$ by using Eqs. (1) and (2). The thus obtained $\beta$ and $\gamma$ are thereafter used to reconstruct a modified Arrott plot. Consequently, new $M_S(T)$ and $\chi_0^{-1}(T)$ are generated from the linear extrapolation in the high field region, and a new set of $\beta$ and $\gamma$ will be acquired. This procedure should be repeated until $\beta$ and $\gamma$ are convergent. The obtained critical exponents from this method are independent on the initial parameters, thus guaranteeing the reliability of the analysis and that the obtained critical exponents are intrinsic. The final modified Arrott plot with $\beta = 0.351(1) \pm 0.001$ and $\gamma = 1.413(5) \pm 0.003$ is presented in Fig. 4. It is clear that the isotherms in the high magnetic field region are a set of parallel straight lines, indicating the reliability of the above analysis. In addition, the final $M_S(T)$ and $\chi_0^{-1}(T)$ with solid fitting curves are depicted in Fig. 5(a), thus the critical exponents $\beta = 0.344(5)$ with $T_C = 273.76(3)$ K and $\gamma = 1.406(1)$ with $T_C = 273.88(4)$ K are obtained.

The Kouvel-Fisher (K-F) method can also be employed to fit the critical exponents and critical temperature, which is expressed as [41]:

$$\frac{M_S(T)}{dM_S(T)/dT} = \frac{T-T_C}{\beta} \qquad (5)$$

$$\text{and} \quad \frac{\chi_0^{-1}(T)}{d\chi_0^{-1}(T)/dT} = \frac{T-T_C}{\gamma}, \qquad (6)$$

where $M_S(T)/(dM_S(T)/dT)$ and $\chi_0^{-1}(T)/(d\chi_0^{-1}(T)/dT)$ are linearly dependent on temperature with the slopes of $1/\beta$ and $1/\gamma$, respectively. As is shown in Fig. 5(b), the linear fits give $\beta = 0.346(4)$ with $T_C = 273.75(7)$ K and $\gamma = 1.364(9)$ with $T_C = 273.97(9)$ K, respectively. Apparently, these obtained critical exponents and critical temperatures are consistent with those obtained from the iterative modified Arrott plot, confirming again the reliability of the above analysis and the intrinsicality of the obtained parameters.

The iterative modified Arrott plot gives the critical exponents $\beta$ and $\gamma$, while the critical exponent $\delta$ can be obtained by using Eq. (3). Fig. 6 shows the isothermal magnetization $M(H)$ at a critical temperature $T_C = 274$ K and the inset shows the plot at a log-log scale. According to Eq. (3), the $M(H)$ at $T_C$ should be a straight line in the log-log scale with the slope of $1/\delta$, thus giving $\delta = 5.02(1)$. To check the reliability of such analysis, $\delta$ was also calculated by using the Widom scaling relation [42]:

$$\delta = 1 + \frac{\gamma}{\beta}, \qquad (7)$$

which gives $\delta = 5.02(6)$ and $\delta = 4.94(0)$ by using $\beta$ and $\gamma$ obtained with modified Arrott plot and Kouvel-Fisher plot, respectively, which are consistent with those fitted by using Eq.(3).

From above analysis, a set of critical exponents are obtained, which are actually self consistent. It is of essential importance to check whether the obtained critical exponents and $T_C$ can generate a scaling equation of state for $Fe_5GeTe_2$, i.e., to examine the reliability of these critical exponents by using the scaling analysis. According to the scaling hypothesis, for a magnetic system in the critical asymptotic region, the scaling equation of state can be expressed as [43]:

$$M(H, \varepsilon) = \varepsilon^\beta f_\pm(\frac{H}{\varepsilon^{\beta+\gamma}}) \quad (8),$$

where $M(H, \varepsilon)$, $H$, and $T$ are variables; $f_+$ for $T > T_C$ and $f_-$ for $T < T_C$ are the regular functions. Through the renormalization process, Eq. (8) can also be written as:

$$m = f_\pm(h), \quad (9)$$

where $m \equiv \varepsilon^{-\beta} M(H, \varepsilon)$ is the renormalized magnetization and $h \equiv \varepsilon^{-(\beta+\gamma)}$ is the renormalized field. If the critical exponents $\beta$, $\gamma$ and $\delta$ could be properly chosen, the scaled $m(h)$ plot will fall onto two universal curves: one for $T > T_C$ and the other one for $T < T_C$, indicating that the interactions are properly renormalized in the critical regime following the scaling equation of state. The scaled $m$ and $h$ curves are plotted in Fig. 7(a), which clearly show two branches below and above $T_C$, indicating that the obtained critical exponents are reliable. Inset of Fig. 7(a) is the same data while is plotted in a log-log form, which can show the two branches more clearly. To support the analysis, we used a more rigorous method by plotting $m^2$ against $h/m$, seen in Fig. 7(b) in which all data apparently separate into two curves below and above $T_C$. The reliability of the obtained critical exponents and $T_C$ can also be examined by checking the scaling of the magnetization curves. The scaling state equation of magnetic systems is [43]:

$$\frac{H}{M^{\delta}} = h(\frac{\varepsilon}{H^{1/\beta}}),  \qquad (10)$$

where $h(x)$ is a scaling function. From Eq. (10), the curves of $\varepsilon H^{-(\beta\delta)}$ vs. $MH^{-1/\delta}$ should fall on one universal curve [44], as seen by the inset of Fig. 7(b). The $T_C$ lies on the zero point of $\varepsilon H^{-(\beta\delta)}$ axis. As a result, the well rescaled curves further confirm that the obtained critical exponents and $T_C$ are reliable and consistent with the scaling hypothesis.

It is valuable to compare the critical exponents of $Fe_5GeTe_2$ with those of other layered vdW magnets and those predicted by various models. The critical exponents of $Fe_5GeTe_2$ obtained by using different analysis techniques and different theoretical models are summarized in Table I, together with those of other several FM vdW magnets including $Fe_{3-x}GeTe_2$ ($x$ = 0, 0.15, and 0.36), $Cr_2Si_2Te_6$, and $Cr_2Ge_2Te_6$. The previous comprehensive study reached a conclusion that the critical exponent $\beta$ for a 2D magnets lies in the range of ~ $0.1 \leq \beta \leq 0.25$ [45]. It is apparent that the $\beta$ values of $Cr_2Si_2Te_6$ and $Cr_2Ge_2Te_6$, which were verified as 2D Ising magnets [46, 47], are actually within the window, while those of $Fe_{3-x}GeTe_2$ and $Fe_5GeTe_2$ are apparently larger than 0.25, thus excluding the 2D Ising model [29, 30]. Moreover, the $\gamma$ values of $Fe_{3-x}GeTe_2$ and $Fe_5GeTe_2$ are much larger than those for the tricritical mean-field and 3D Ising models [39, 40], suggesting the two models are not appropriate. Combining the $\beta$ and $\gamma$ values, the magnetic critical behavior in $Fe_5GeTe_2$ should have a 3D nature, indicating that the interlayer magnetic exchange can not be neglected. The difference of the magnetic characteristics between $Fe_{3-x}GeTe_2$ and $Cr_2(Si,Ge)_2Te_6$ is attributed to the smaller vdW gaps in $Fe_{3-x}GeTe_2$ [17], which are capable of giving rise to stronger interlayer magnetic exchange. It is naturally a hypothesis that the vdW gap in $Fe_5GeTe_2$ should hold the same situation as that in $Fe_{3-x}GeTe_2$. To achieve more insights, the critical exponents of $Fe_5GeTe_2$ should be compared with the several 3D models more carefully. The $\beta$ of $Fe_5GeTe_2$ is much closer to that of the 3D XY model [40] while the $\gamma$ is closer to that of the 3D Heisenberg model [39]. The fact indicates that the obtained critical exponents of $Fe_5GeTe_2$ can not be simply categorized into any conventional universality classes.

For a homogenous magnet, the universality class of the magnetic phase transition depends on the magnetic exchange distance $J(r)$. According to the renormalization group theory analysis, the magnetic exchange decays with the distance $r$ in a form $J(r) \sim e^{-r/b}$ for the short-range magnetic exchange and $J(r) \sim r^{-(d+\sigma)}$ for the long-range exchange, where $r$ is the exchange distance, $b$ is the spatial scaling factor, $d$ is the dimensionality of the system, and the positive constant $\sigma$ denotes the range of exchange interaction [48, 49]. Moreover, within this theory model the susceptibility exponent $\gamma$ is defined as [48]:

$$\gamma = 1 + \frac{4}{d}\left(\frac{n+2}{n+8}\right)\Delta\sigma + \frac{8(n+2)(n-4)}{d^2(n+8)^2}\left[1 + \frac{2G(\frac{d}{2})(7n+20)}{(n-4)(n+8)}\right]\Delta\sigma^2, \qquad (11)$$

where $n$ is the spin dimensionality and $\Delta\sigma = (\sigma - d/2)$ and $G\left(\frac{d}{2}\right) = 3 - \frac{1}{4}\left(\frac{d}{2}\right)^2$. For 3D materials ($d = 3$) with $3/2 \leq \sigma \leq 2$, the magnetic exchange decays relatively slowly as $J(r) \sim r^{-(d+\sigma)}$ due to a long-range magnetic exchange. For $\sigma > 2$, the 3D Heisenberg model is valid for 3D isotropic magnet, where $J(r)$ decreases faster than $r^{-5}$ due to the short-range magnetic exchange, while when $\sigma \leq 3/2$, the mean-field model works and $J(r)$ decreases slower than $r^{-4.5}$ [48, 49]. To obtain the values of $d$, $n$, and $\sigma$ for $Fe_5GeTe_2$, a method similar to that in Ref. [49] was adopted. In this method, $\sigma$ is adjusted in Eq. (11) with several sets of $\{d : n\}$ to get a proper $\gamma$ that is close to the experimental value ($\sim 1.364$). The obtained $\sigma$ is then used to calculate other critical exponents including $\beta$, $\delta$, $\nu$ and $\alpha$ by the following equations: $\nu = \gamma/\sigma$, $\alpha = 2 - \nu d$, $\beta = (2 - \alpha - \gamma)$, and $\delta = 1 + \gamma/\beta$. After repeating the calculations for several sets of $\{d : n\}$, with the typical results being summarized Table II, it is found that $\{d : n\} = \{3 : 3\}$ and $\sigma = 1.916$ yielded critical exponents of $\beta = 0.3851$, $\gamma = 1.3613$ and $\delta = 4.5351$ match well with the experimental values. Such a result indicates the 3D Heisenberg type magnetic exchange in $Fe_5GeTe_2$ with long-range interaction decaying as $J(r) \approx r^{-4.916}$, which is consistent with our analysis presented above.

The magnetic exchange models for qiasi-2D vdW magnets have been an immensely investigated subject. For $Cr_2Si_2Te_6$, the magnetic critical behavior analysis and neutron scattering studies consistently suggest the universality class of 2D Ising model for its magnetic critical behavior [26, 50]. Because $Cr_2Ge_2Te_6$ has an enhanced interlayer exchange due to the smaller vdW gap and larger cleavage energy than $Cr_2Si_2Te_6$, its critical behavior shows a transition from the 2D Ising-type to a 3D tricritical mean-field type [51]. The mean distance between the two adjacent Te layers that across the vdW gap in $Fe_3GeTe_2$ is 0.423 nm [52], which is rather close to that of $CrGeTe_3$, 0.377 nm [50], which presumably can account for the 3D Heisenberg characteristics of the critical behavior. Previous studies on $Fe_5GeTe_2$ indicate small magnetic anisotropy at high temperature [20], so the 3D magnetism for the critical behavior in $Fe_5GeTe_2$ is reasonable. Moreover, it is found that the magnetic anisotropy in $Fe_{3-x}GeTe_2$ strongly depends on the Fe deficiency [53], which can be largely suppressed with increasing the deficiency content $x$. If we pay a close attention to the critical exponents of $Fe_5GeTe_2$, it is easily found that they are much closer to those of Fe deficient $Fe_{3-x}GeTe_2$ [29], likely further demonstrating the weak magnetic anisotropy in $Fe_5GeTe_2$. However, the possible transition between different universality classes of models of the critical behavior should be carefully checked, if we recall into our mind that earlier neutron measurements on $NiPS_3$ actually unveiled a critical phase transition between 3D and 2D at the temperature of ~ $0.9T_C$ [54]. The critical phase transition between the 2D anisotropic Heisenberg model and the 3D magnetism below $T_C$ was also observed in $MnPS_3$ [55]. Though such possibility has not been examined yet in $Fe_{3-x}GeTe_2$, considering that $Fe_{3-x}GeTe_2$ indeed shares similarities as $MPS_3$ (MM = Mn, Fe, and Ni) in that they all have 2D antiferromagnetic ground state with the ferromagnetic layers in them order antiferromagnetically along the $c$-axis at low temperature, as well as the 3D critical behavior near $T_C$, the critical phase transition definitely need to be checked in $Fe_{3-x}GeTe_2$. For $Fe_5GeTe_2$, it is somewhat different from $MPS_3$ and $Fe_{3-x}GeTe_2$, which behaves as an easy-axis vdW ferromagnet with the magnetic moments preferring to align along the $c$-axis but with weak anisotropy at high temperature due to the easy polarization of moments and the interaction between the FM layers is still FM. However, the magnetism of $Fe_5GeTe_2$ is somewhat complex due to the multiple Fe sublattices and composition tunable $T_C$. It is revealed that the magnetic moments on Fe(1) sublattice order below ~ 100 - 120 K while the majority of the moments order at $T_C$ [21]. Short-range order associated with occupations of split sites of Fe(1) is also present. Additionally, the magnetic anisotropy is enhanced at low temperature. Regarding these, more studies to establish the precise spin structure at low temperature are extremely desired.

## IV. CONCLUSION

In summary, we have investigated the magnetic critical behavior in vicinity of the PM to FM phase transition in the quasi-2D van der Waals ferromagnet $Fe_5GeTe_2$ which has a near room temperature $T_C$ of approximately 270 K. The estimated critical exponents $\alpha$, $\beta$ and $\gamma$ values from the various techniques and theoretical models show nice consistence with each other and follow the scaling behavior well. The critical exponents suggest a second order phase transition and they do not belong to any single universality class of model, just lying between the 3D Heisenberg model and the 3D XY model. The magnetic exchange distance is found to decay as $J(r) \approx r^{-4.916}$, which is close to that of 3D Heisenberg model with long-range exchange. The critical phenomena indicate weak magnetic anisotropy of $Fe_5GeTe_2$ at high temperature, possibly due to its small vdW gap. The very recent calculations indicate that monolayer formation energy of $Fe_5GeTe_2$ lies inside the energy range of other 2D materials [56], and the synthesis of the monolayer is therefore highly expected. Moreover, considering the tunable $T_C$ which can even to be ~ 350 K [20, 21, 57], the investigation on the precise magnetic structure of $Fe_5GeTe_2$ would find extraordinary opportunities for applications in next-generation spintronic devices.

## ACKNOWLEDEMENTS

The authors acknowledge the support by the Natural Science Foundation of Shanghai (Grant No. 17ZR1443300) and the National Natural Science Foundation of China (Grant No. 11874264). Y.F.G. acknowledges the starting grant of ShanghaiTech University and the Program for Professor of Special Appointment (Shanghai Eastern Scholar). L.M.C. is supported by the Key Scientific Research Projects of Higher Institutions in Henan Province (19A140018). The authors also thank the support from the Analytical Instrumentation Center (#SPST-AIC10112914), SPST, ShanghaiTech University.

**Table I.** A summary of the critical exponents of $Fe_5GeTe_2$, $Fe_{3-x}GeTe_2$, $Cr_2Si_2Te_6$, $Cr_2Ge_2Te_6$ and those predicted by different models (MAP: Modified Arrott plot; KF: Kouvel-Fisher method; CI: critical isotherm analysis).

| Composition | Reference | Technique | $\beta$ | $\gamma$ | $\delta$ | {d:n} | J(r) |
|---|---|---|---|---|---|---|---|
| $Fe_5GeTe_2$ | This work | MAP | 0.351(1) | 1.413(5) | 5.02(6) | {3:3} | $r^{-4.916}$ |
|  | This work | KF | 0.346(4) | 1.364(9) | 4.94(0) |  |  |
|  | This work | CI |  |  | 5.02(1) |  |  |
| 3D Heisenberg | [4] | Theory | 0.365 | 1.386 | 4.8 |  |  |
| 3D XY | [4] | Theory | 0.345 | 1.316 | 4.81 |  |  |
| 3D Ising | [4] | Theory | 0.325 | 1.24 | 4.82 |  |  |
| Tricritical mean field | [5] | Theory | 0.25 | 1.0 | 5 |  |  |
| Mean field | [4] | Theory | 0.5 | 1.0 | 3 |  |  |
| $Fe_{2.64}Ge_{0.87}Te_2$ | [12] | KF | 0.372(4) | 1.265(1) | 4.401(6) | {3:3} | $r^{-4.89}$ |
| $Fe_{2.85}GeTe_2$ | [13] | KF | 0.363 | 1.228 | 4.398 |  | $r^{-4.8}$ |
| $Fe_3GeTe_2$ | [3] | KF | 0.322(4) | 1.063(8) | 4.301(6) |  | $r^{-4.6}$ |
| $Cr_2Si_2Te_6$ | [14] | KF | 0.175(9) | 1.562(9) | 9.925(5) | {2:1} | $r^{-3.63}$ |
| $Cr_2Ge_2Te_6$ | [15] | KF | 0.200(3) | 1.28(3) | 7.405 | {2:1} | $r^{-3.52}$ |

**Table II.** Critical exponents calculated by the renormalization group theory.

| d | n | $\sigma$ | $\beta$ | $\gamma$ | $\delta$ |
|---|---|---|---|---|---|
| 3 | 3 | 1.9160 | 0.3851 | 1.3613 | 4.5351 |
| 2 | 1 | 1.3603 | 0.3168 | 1.370 | 5.3241 |
| 2 | 3 | 1.2740 | 0.3904 | 1.370 | 4.5096 |

**Figure captions:**

**Fig. 1.** (a-b) Crystal structure of $Fe_5GeTe_2$ viewed from directions along the *a*- and *c*-axis, respectively. (c) Image of a typical $Fe_5GeTe_2$ single crystal. (d-e) Single crystal X-ray diffraction patterns in the reciprocal space along the (*0 k l*), (*h 0 l*), and (*h k 0*) directions.

**Fig. 2.** (a) Temperature dependence of magnetization $M(T)$ for $Fe_5GeTe_2$ under $H = 1$ kOe. The inset shows the inverse susceptibility plotted against temperature and the straight dotted line is Curie-Weiss law fitting. (b) Isothermal magnetization $M(H)$ measured at 2 K. (c) Typical initial magnetization $M(H)$ curves measured from 261 K to 285 K with an interval of 1 K. (d) Arrott plots in the form of $M^2$ vs $H/M$ (mean

field model) around $T_C$.

**Fig. 3.** The isotherms of $M^{1/\beta}$ versus $(H/M)^{1/\gamma}$ with (a) 3D Heisenberg model, (b) 3D Ising model, (c) 3D XY model, (d) Tricritical mean-field model and (e) 2D Ising model. (f) Normalized slope versus temperature curves for six sets of critical exponents.

**Fig. 4.** Modified Arrott Plot of isotherms with $\beta = 0.351(1) \pm 0.001$ and $\gamma = 1.413(5) \pm 0.003$ for $Fe_5GeTe_2$.

**Fig. 5.** (a) Temperature dependence of the spontaneous magnetization $M_S$ (left) and the inverse initial susceptibility $\chi_0^{-1}(T)$ (right) with solid fitting curves for $Fe_5GeTe_2$. (b) Kouvel-Fisher plots of $M_S(T)/(dM_S(T)/dT)$ (left) and $\chi_0^{-1}(T)/(d\chi_0^{-1}(T)/dT)$ (right) with solid fitting curves for $Fe_5GeTe_2$.

**Fig. 6.** Isotherm $M(H)$ collected at $T_C = 274$ K for $Fe_5GeTe_2$. Inset: the same plot in log-log scale with a solid fitting curve.

**Fig. 7.** (a) Renormalized magnetization $m \equiv \varepsilon^{-\beta} M(H, \varepsilon)$ as a function of renormalized field $h \equiv \varepsilon^{-(\beta+\gamma)}$ below and above $T_C$ for $Fe_5GeTe_2$. Inset is the same $m(h)$ data in log-log scale. (b) Plot in the form of $m^2(h/m)$ for $Fe_5GeTe_2$. Inset shows the plot of $\varepsilon H^{-(\beta\delta)}$ vs. $MH^{-1/\delta}$ below and above $T_C$.

**Figure 1**

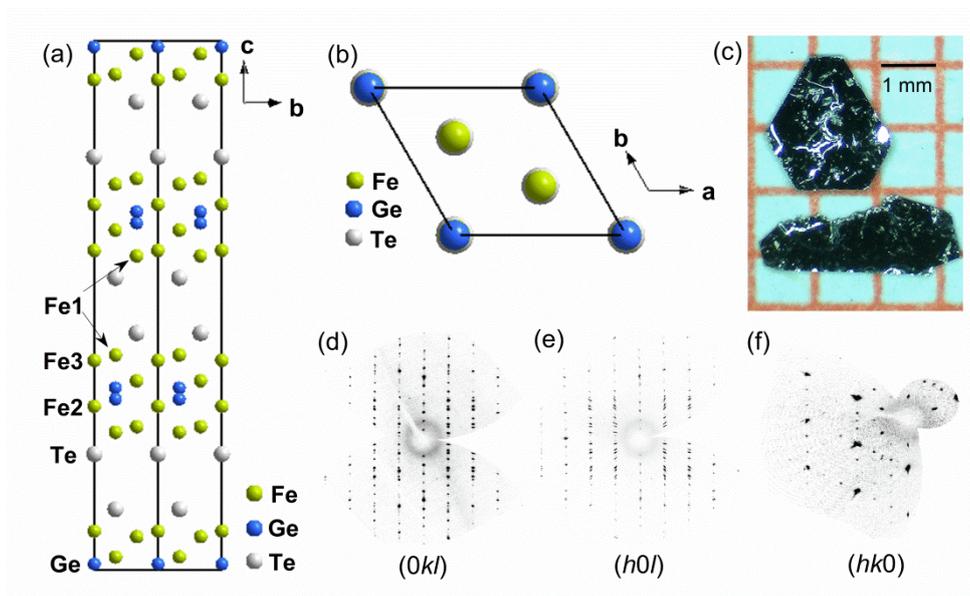

**Figure 2**

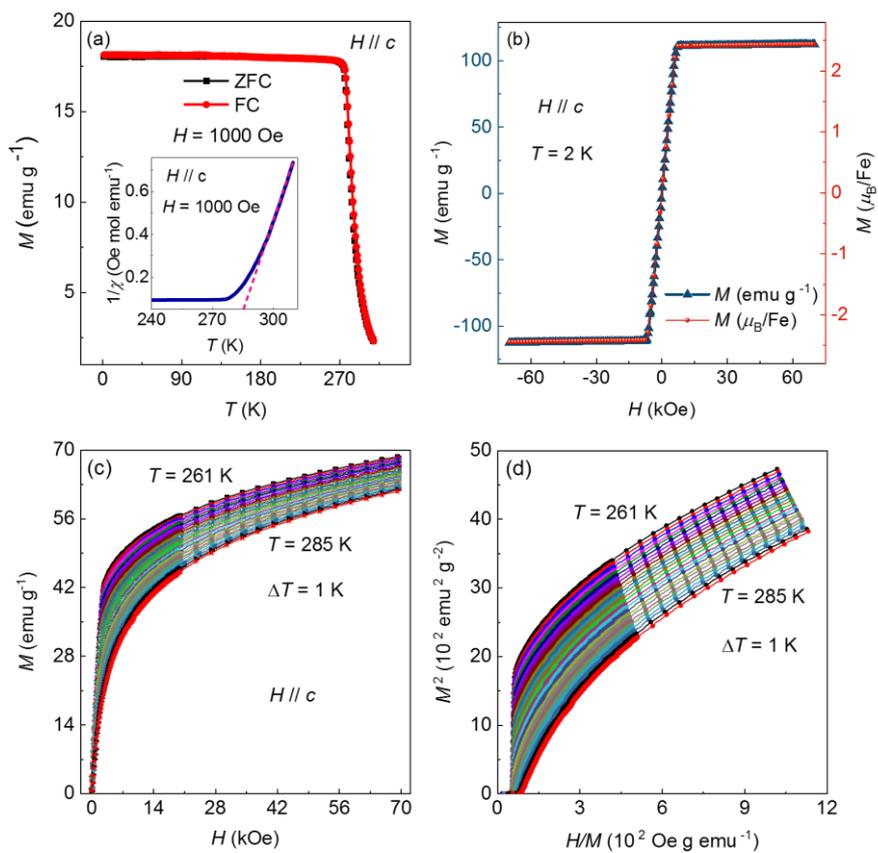

**Figure 3**

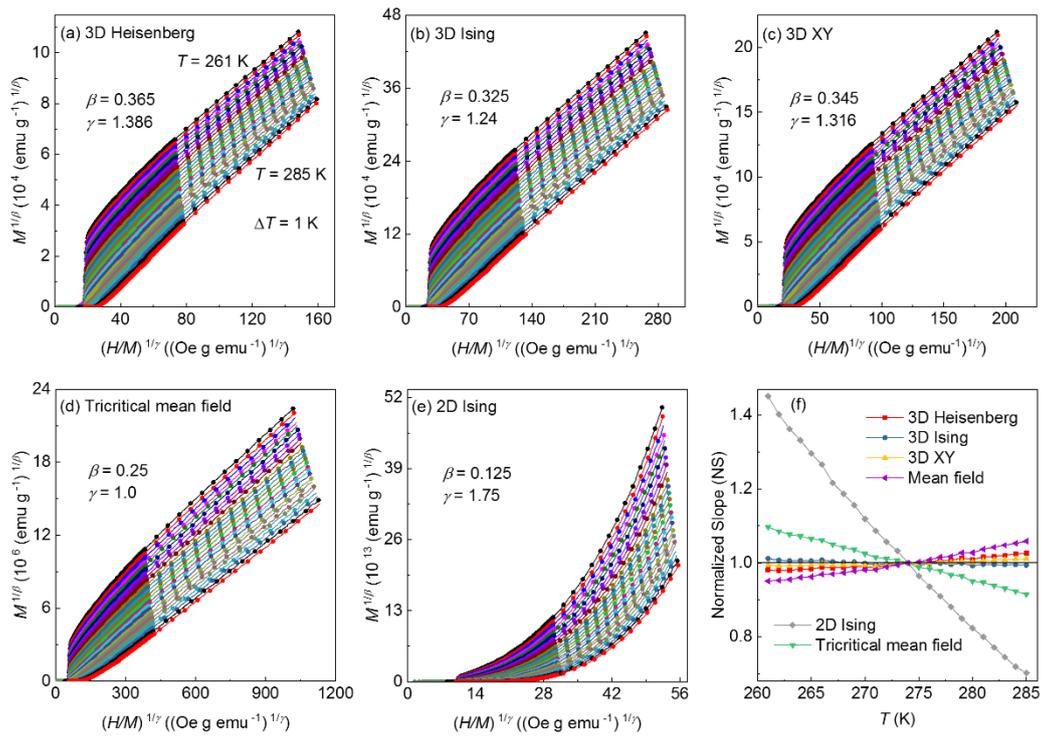

**Figure 4**

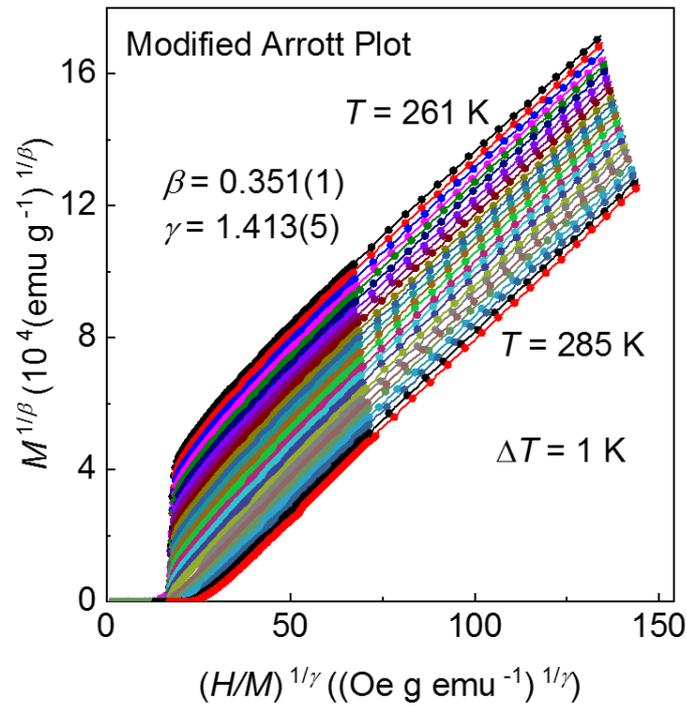

**Figure 5**

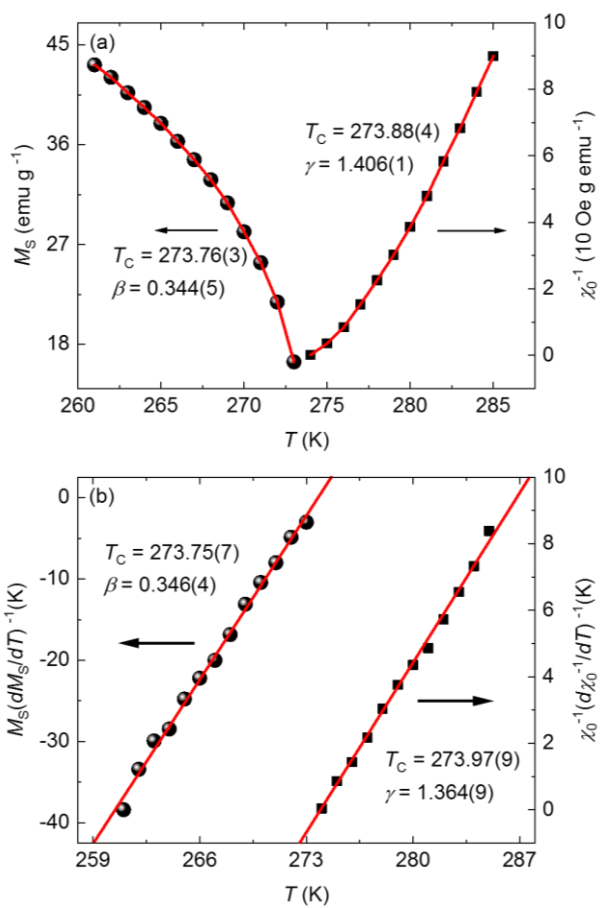

**Figure 6**

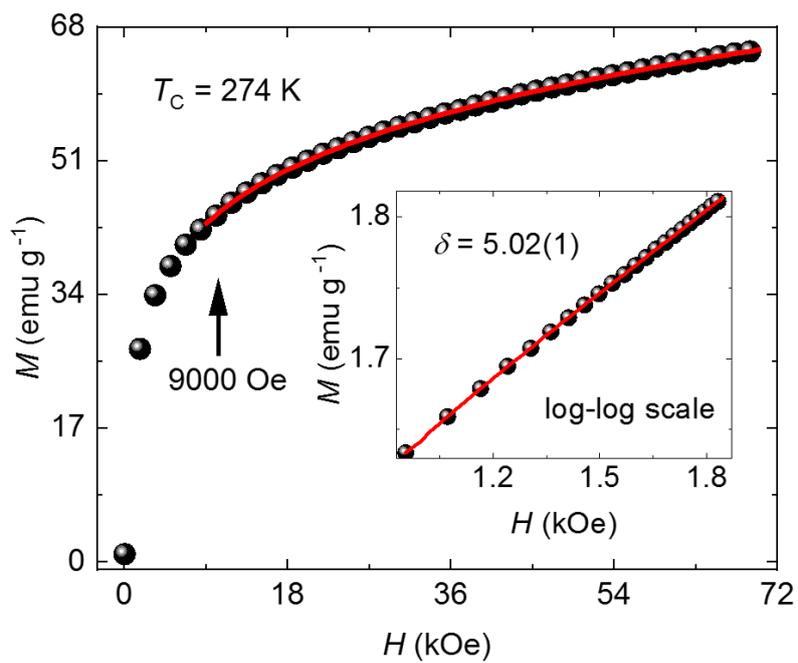

**Figure 7**

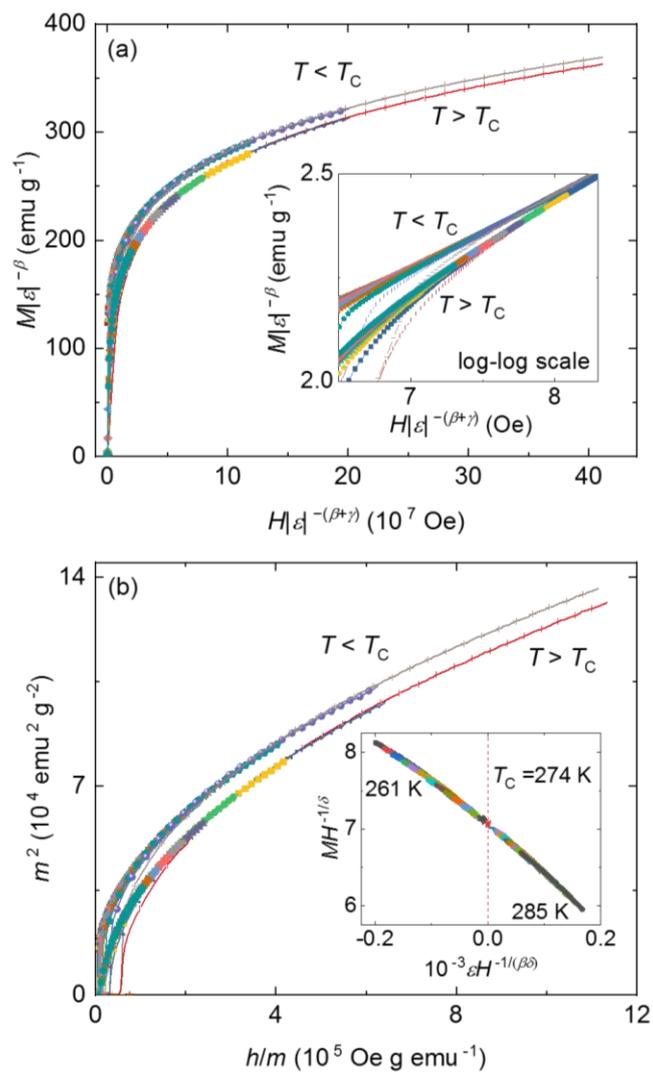